\documentclass[nofootinbib,
reprint
]{revtex4}

\usepackage{graphicx}
\usepackage{mathrsfs}
\usepackage{yfonts}
\usepackage{tipa}
\usepackage{bm}
\usepackage{bbm}
\usepackage{amsmath}
\usepackage{amssymb}
\usepackage{amsthm}
\usepackage{hyperref}
\usepackage{amsfonts}
\usepackage{mathtools}
\usepackage{latexsym}
\usepackage{relsize}
\usepackage[greek,english]{babel}
\usepackage{booktabs}
\usepackage[iso-8859-7]{inputenc}
\usepackage{tikz}
\usepackage{wrapfig}
\usepackage{subfigure}

\begin{document}

\title{Noncommutative black holes in extended anti-de Sitter phase space}

\author{\textbf{Athanasios G. Tzikas}}
\email{athanasios.tzikas@unibg.it} 
\affiliation{Universit\`a degli Studi di Bergamo, Dipartimento di Ingegneria e Scienze
Applicate, Viale Marconi 5, 24044 Dalmine (Bergamo) Italy.}

\begin{abstract}
We study  thermodynamic aspects  of ordinary and lower dimensional noncommutative black holes within an extended anti-de Sitter  phase space by treating the negative cosmological constant and the minimal cut-off length as  thermodynamic variables representing the  pressure and tension of the system, respectively.  In four-dimensional spacetime, the regular black hole exhibits a small/large black hole phase transition analogous to the liquid/gas transition of a Van der Waals gas. The three-dimensional case  demonstrates global and local thermodynamic stability, while the two-dimensional case reveals a novel type of transition referred to as the anti-Hawking-Page transition.
\end{abstract}

\maketitle

\section{Introduction}
\label{intro}

The breakdown of general relativity at the ``heart" of a black hole has driven  an ongoing pursuit to reconcile quantum mechanics with gravity over the past century (see \cite{GiH93,Rov00,GSW12} for reviews). Despite significant progress, a complete formulation of quantum gravity in $(3+1)$-dimensions remains incomplete. This challenge arises from the dimensional nature of Newton's constant, which makes gravity a non-renormalizable field theory \cite{Sho07}. Therefore, rather than pursuing a complete quantum gravitational theory, short-distance pathologies can be addressed through effective approaches. These include noncommutative geometry \cite{Nic09}, loop quantum gravity \cite{Mod05}, the generalized uncertainty principle \cite{IMN13}, nonlocal gravity \cite{MMN11}, and asymptotically safe gravity \cite{LaR05} (see also references in \cite{Nic18}).

During this work, we focus on black holes arising from the noncommutativity of spacetime \cite{NSS06}. These objects, known as noncommutative black holes (NCBHs), represent a class of black hole solutions to the Einstein field equations that incorporate quantum mechanical effects  in the short distance regime of the gravitational field of a black hole. NCBHs are one of the richest classes of non-singular (or regular) black holes, which include dirty \cite{NiS10}, charged \cite{ANS++07}, rotating \cite{SmS10}, charged and rotating \cite{MoN+10}, (anti-)de Sitter \cite{NiT11,MaN11}, and higher or lower-dimensional cases \cite{Riz06,SSN09,MuN11,RKB++13,GNS21}. A defining characteristic of noncommutative geometry is the presence of quantum fluctuations, which address short-distance singularities. These fluctuations are encoded within the commutator
\begin{equation} \label{commut}
\left[ x^{\mu},x^{\nu} \right]= i  \theta^{\mu\nu} ,
\end{equation}
where $\theta^{\mu\nu}$ is an anti-symmetric tensor with dimensions of  length squared that discretizes the spacetime vectors $x^{\mu}$, much like the Planck constant $\hbar$ discretizes the phase space. Eq.~\eqref{commut} indicates the ``fuzziness" of spacetime at short distances through the uncertainty of  spacetime itself and implies that ``points" no longer exist.  
On physical grounds, we can think of $\sqrt{\theta}$ as the minimal cut-off length, very close to the Planck length, under which gravity cannot be tested. A simple way to address this issue at the center of a black hole is to impose that its mass is smeared throughout the center, just like a Gaussian distribution of minimal width $\sqrt{\theta}\,$, instead of being a vacuum Dirac delta function. Hence, the black hole can be described by a source of anisotropic fluid with energy density of the form \cite{SSN09}
\begin{equation} \label{NVdensity}
\rho(r) = \frac{M}{(4\pi \theta)^{d/2}} \exp({-r^2/4\theta}) \,,
\end{equation}
where $d$ indicates the spatial dimensionality of the manifold, and $M$ is the black hole mass which is diffused throughout a region of linear size $\sqrt{\theta}\,$. NCBHs match known black hole solutions at large distances  while introducing new physics at short scales. A key result is that, for all NCBHs, the curvature singularity at the origin is smeared out  and replaced by a regular core. 
Another significant feature is the improvement of thermodynamics, with the Hawking temperature reaching a maximum before running a positive heat capacity, ultimately leading to a zero-temperature remnant configuration.

We now turn to another research area, known as black hole chemistry \cite{KuM14,KuM17}. The central idea is to treat the cosmological constant $\Lambda$ in an anti-de Sitter (AdS) background as a thermodynamic variable, which is associated with the positive pressure of the system. Along this line of reasoning, we extend further our phase space by also treating the noncommutative minimal length $\sqrt{\theta}$ as a  thermodynamic variable. Consequently,  an additional negative pressure$-$referred to as noncommutative pressure$-$emerges and serves as a tension that counteracts gravitational collapse near the center \cite{MiX17, Lia17}. The framework of chemistry improves black hole thermodynamics and reveals new  properties, analogous to established chemical phenomena  \cite{KuM12,Dol12,AKM13,AKM+14,RKM14,HKM14,FKM14, FMM15,Tzi19}. The basic features of black hole chemistry are summarized below:
\begin{itemize}
\item The cosmological constant $\Lambda$ and the minimal length $\sqrt{\theta}$ are connected to the  thermodynamic pressure $P$  and the noncommutative tension $P_{\theta}$ through the relations
\begin{equation} \label{press}
P=-\frac{\Lambda}{8\pi}=\frac{d (d-1)}{16\pi l^2} \quad \mathrm{and} \quad P_{\theta} = -\frac{d (d-1)}{16\pi \theta}
\end{equation}
where $l$ is the  AdS radius of the cosmological background.
\item The  black hole mass  no longer represents the internal energy $U$. Instead, it represents the chemical enthalpy $H$ of the system \cite{KRT09}.
\item The conjugate quantities of $P$ and $P_{\theta}$ are given respectively by
\begin{equation} \label{volume}
V=\left( \frac{\partial M}{\partial P} \right) _{S,P_{\theta}} \quad \mathrm{and} \quad V_{\theta}=\left( \frac{\partial M}{\partial P} \right) _{S,P} 
\end{equation}
where $S$ is the black hole entropy. The quantity $V$ represents the thermodynamic volume, whereas  $V_{\theta}\,$, which also has dimensions of volume, is referred to as the noncommutative volume \cite{MiX17}.
\item The first law of thermodynamics is expressed as
\begin{equation} \label{1stLaw}
\mathrm{d} M = T \mathrm{d} S + V \mathrm{d} P + V_{\theta} \mathrm{d} P_{\theta} \,.
\end{equation}
\item  The thermodynamic potential  governing our system is the Gibbs free energy $G\,$, given by
\begin{equation}
G=M-T S \,.
\end{equation}
It provides insight into the global thermodynamic stability of the system, revealing possible first- or second-order phase transitions. The Gibbs energy can also be obtained from the instanton gravitational action $\mathcal{I}$ (see \cite{KuM12} for example). This Euclidean action is related to the Gibbs energy through the expression $G = \mathcal{I}/\beta$, where $\beta$ represents the period of the imaginary time, given by the inverse of the Hawking temperature, i.e., $\beta = 1/T$.
\item Having defined the pressures and volume of the black hole, we can now distinguish between the two specific heats: one at constant pressure $C_P$, and one at constant volume $C_V$, expressed as
\begin{equation} \label{specific_heat}
C_P = T \left( \frac{\partial S}{\partial T} \right)_{P} \quad \text{and} \quad C_V = T \left( \frac{\partial S}{\partial T} \right)_{V}.
\end{equation}
These specific heats characterize the local thermodynamic stability of the system, with divergences indicating potential phase transition points for the black hole.
\end{itemize}
Our goal is to extend the works of \cite{HaP83, FMM15} by examining the chemical properties of ordinary and lower-dimensional NCBHs within the framework of noncommutative geometry. The paper is organized as follows: in Sec.~\ref{4dNC_chem}, we review the solution of the noncommutative Schwarzschild-anti-de Sitter  black hole in $(3+1)$-dimensions and  study its chemical properties.  We demonstrate that the black hole can undergo a first-order phase transition between small/large stable black hole, analogous to the liquid/gas coexistence in a Van der Waals gas. Moreover, unlike the non-extended phase space, the extended phase space enhances the analogy between the thermodynamic variables of the black hole equation of state and those of the Van der Waals equation. 
In Sec.~\ref{3dNC_chem}, we derive the line element for the noncommutative BTZ (NCBTZ) black hole, which is a solution arising from the noncommutativity of a $(2+1)$-dimensional manifold. The analysis of its thermodynamic properties in the extended phase space indicates both local and global thermodynamic stability, with no indication of a potential phase transition. In Sec.~\ref{2dNC_chem}, we derive the solution for a noncommutative anti-de Sitter black hole in $(1+1)$-dimensions and repeat the same procedure to examine its chemistry. We find that the $(1+1)$-dimensional regular black hole exhibits a reverse Hawking-Page transition between pure AdS radiation and a stable black hole, in contrast to its singular counterpart, which remains thermodynamically stable at all times. Finally, in Sec.~\ref{concl}, we summarize our results.

\section{Noncommutative black hole chemistry in ($3+1$)-dimensions} 
\label{4dNC_chem}

\subsection{The line element} 
 
\begin{figure}[h!] 
\includegraphics[width=0.53 \textwidth]{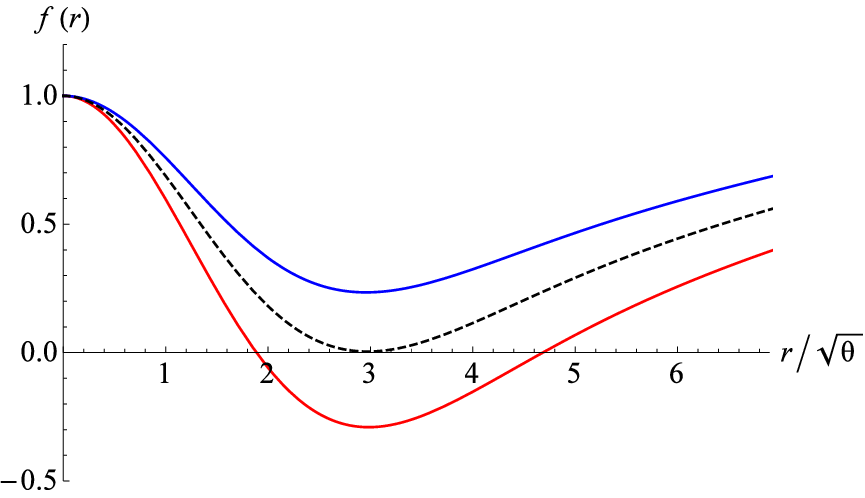}
\caption{We  set $l=20\sqrt{\theta}$ and we plot the   metric potential \eqref{4dmetric}    for $M/l=0.075$ (blue solid curve), for $M_0/l\approx 0.097$ (black dashed curve) and for $M/l=0.125$ (red solid curve).}
\label{fig:4dNCmetric}
\end{figure}

We review the line element of a static, neutral, spherically symmetric, noncommutative diffused gravitational source of mass $M$ inside an AdS background $(\Lambda<0)$ with  energy density \cite{NiT11}
\begin{equation} \label{4D_density}
\rho(r) = \frac{M}{(4\pi \theta)^{3/2}} \exp({-r^2/4\theta}) \,.
\end{equation}
In order to derive the  line element, we have to solve the Einstein field equations with a stress-energy tensor $\cal{T}^{\mu}_{\nu}$ resulting from the above density-profile. The components of $\cal{T}^{\mu}_{\nu}$ are specified through the covariant conservation condition $\nabla_{\mu} \mathcal{T}^{\mu}_{\nu}=0$ along with the Schwarzschild-like property $g_{tt}=-1/g_{rr}\,$, which is implied through the relation $\mathcal{T}^t_t=\mathcal{T}^r_r \rightarrow G^t_t=G^r_r$ \cite{Jac07}. The tensor turns out to have the diagonal form of
\begin{equation} 
\mathcal{T}^{\mu}_{\nu}=\mathrm{diag} \left( -\rho(r), \ p_{\mathrm{r}}(r), \ p_{\perp}(r), \ p_{\perp}(r) \right) .
\end{equation}
We notice that there are two non-vanishing pressure terms; the radial pressure $p_{\mathrm{r}}(r)$ and the tangential pressure $p_{\perp}(r)\,$,  with $p_{\mathrm{r}}(r) \neq p_{\perp}(r)\,$, corresponding to the case of an anisotropic fluid. On physical grounds, a non-vanishing radial pressure is necessary to counterbalance the inward gravitational pull, preventing the source from collapsing into a singular matter point. The two pressures satisfy the following relations:
\begin{equation}
p_{\mathrm{r}} (r)=-\rho(r) \qquad \mathrm{and} \qquad  p_{\perp}(r)= - \rho(r) - \frac{r}{2} \frac{\partial \rho(r)}{\partial r}  \,.
\end{equation}
In the presence of a negative cosmological term $\Lambda=-3/l^2$ and in units where $c=\hbar=G=k_{\rm B}=1\,$, we consider the Einstein field equations
\begin{equation} \label{einstein_eq}
R^{\mu}_{\nu}-\frac{R}{2}\delta^{\mu}_{\nu}+\Lambda \delta^{\mu}_{\nu}=8\pi \mathcal{T}^{\mu}_{\nu}
\end{equation}
with a line element of the form
\begin{equation}
\mathrm{d} s^2 = -f(r) \mathrm{d} t^2 + f(r)^{-1} \mathrm{d} r^2 + r^2 \mathrm{d} \Omega^2 \,
\end{equation}
where $ \mathrm{d} \Omega^2= \mathrm{d} \theta^2+\sin ^2 \theta  \mathrm{d} \phi^2$.
Solving \eqref{einstein_eq} with the above density profile \eqref{4D_density}, we get the following metric potential 
\begin{equation}\label{4dmetric}
f(r) = 1-\frac{4M}{r \sqrt{\pi}} \gamma(3/2 , r^2/4\theta)+\frac{r^2}{l^2} \,
\end{equation}
with $\gamma(3/2, r^2/4\theta)$ being the lower incomplete gamma function given by
\begin{equation}\label{gamma_function}
\gamma(3/2 , r^2/4\theta)= \int \limits ^{r^2/4\theta}_{0} \mathrm{d}z \ z^{1/2} e^{-z}\,.
\end{equation}
In contrast to the usual Schwarzschild-AdS black hole \cite{HaP83}, the  metric potential \eqref{4dmetric} exhibits three distinct horizon structures, as shown in Fig.~\ref{fig:4dNCmetric}:
\begin{itemize}
\item Two different horizons  for $M>M_0$ (red solid curve in Fig.~\ref{fig:4dNCmetric}); one Cauchy (inner) horizon $r_-$ and one event (outer) horizon $r_+\,$.
\item One degenerate horizon $r_0=r_-=r_+$   for $M=M_0$ (black dashed curve in Fig. \ref{fig:4dNCmetric}); $M_0$ is the lowest possible black hole mass corresponding to the extremal case with  zero black hole temperature.
\item No  horizons for $M<M_0$ (blue solid curve in Fig.~\ref{fig:4dNCmetric}).
\end{itemize}
The  mass  $M_0$ can be found by solving numerically the system 
\begin{equation}
f(r)=\frac{\partial f(r)}{\partial r}=0 \,.
\end{equation}
For example,  the extremal value of the mass  given in Fig.~\ref{fig:4dNCmetric},  $M_0 \approx 0.097 l \approx 1.94 \sqrt{\theta}\,$,  corresponds to an extremal radius $r_0 \approx 2.98 \sqrt{\theta}\,$. The presence of the cosmological constant modifies the values of the extremal black hole mass $M_0$ and radius $r_0$ compared to the asymptotically flat extremal noncommutative black hole \cite{NSS06}, resulting in $M_0 > 1.9\sqrt{\theta}$ and $r_0 < 3\sqrt{\theta}\,$. Only for very large values of the AdS radius $(l \rightarrow \infty)$ does the mass $M_0$ approach $1.9\sqrt{\theta}\,$, and the degenerate radius $r_0$ tends to $3\sqrt{\theta}\,$, which correspond to the values of the asymptotically flat extremal case.

We notice that in the classical regime $(r \gg  \sqrt{\theta})$, the metric \eqref{4dmetric} matches the well-known Schwarzschild-AdS solution \cite{HaP83}, while near the origin $(r \ll \sqrt{\theta})$, we obtain a regular core. This can be verified by writing \eqref{gamma_function} as 
\begin{equation}
\gamma(3/2,r^2/4\theta) \  \big|_{r^2 \ll 4\theta} \approx \frac{r^3}{12\theta^{3/2}} \,.
\end{equation}
Thus, near the origin, the metric \eqref{4dmetric} can by approximated as
\begin{equation}
f(r) \approx 1-\frac{\Lambda_\mathrm{eff}}{3} r^2 \,
\end{equation}
where
\begin{equation}
\Lambda_{\mathrm{eff}}= \frac{1}{\sqrt{\pi} } \frac{M}{\theta^{3/2}} -\frac{3}{l^2} \,.
\end{equation}
Depending on the value of $M\,$, we get:
\begin{itemize}
\item a repulsive de Sitter core if $M>\frac{3\theta^{3/2}\sqrt{\pi}}{l^2}\,$ $(\Lambda_{\mathrm{eff}}>0)\,.$ 
\item an attractive anti-de Sitter core if $M<\frac{3\theta^{3/2}\sqrt{\pi}}{l^2}\,$ $(\Lambda_{\mathrm{eff}}<0)\,.$ 
\item a local Minkowski core if $M=\frac{3\theta^{3/2}\sqrt{\pi}}{l^2}\,$ $(\Lambda_{\mathrm{eff}}=0)\,.$
\end{itemize}
However, the condition $M \geq M_0 \geq \frac{3\theta^{3/2}\sqrt{\pi}}{l^2}$ must be satisfied for horizons to exist \cite{NiT11}, thereby excluding the AdS core. 

\subsection{The chemistry}

\begin{figure}[h!] 
\includegraphics[width=0.53 \textwidth]{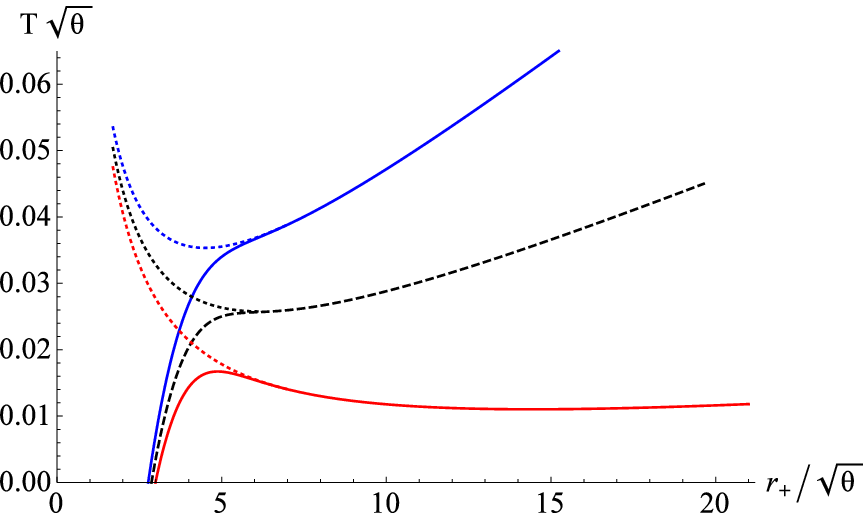}
\caption{The temperature   of a ($3+1$)-dimensional noncommutative black hole is displayed  for $l=18 \sqrt{\theta}\,$ (red solid curve), for $l = l_{\mathrm{c}} \approx 11.15 \sqrt{\theta}\,$ (black dashed curve) and for $l=8 \sqrt{\theta}\,$ (blue solid curve). The dotted curves represent the conventional Hawking-Page temperatures \cite{HaP83}.}
\label{fig:4dNCT}
\end{figure}

We assume that  the known thermodynamic relations can be extended to the case in which a specific microscopic structure of the quantum spacetime is prescribed (see  \cite{NiT11,Nic10} for further discussion).
First, we derive  all the desired thermodynamic quantities for the  metric \eqref{4dmetric}  with respect to the event horizon $r_+\,.$ For simplicity, we set 
\begin{equation}
\gamma(r_+)=\gamma(3/2,r_+^2/4\theta) \qquad \mathrm{and} \qquad  \gamma'(r_+)=\frac{\mathrm{d} \gamma(r)}{\mathrm{d} r} \Big|_{r_+}=\frac{r_+^2}{4\theta^{3/2}} e^{-r_+^2/4\theta} .
\end{equation}
From the horizon condition $f(r_+)=0\,$, we find that the black hole enthalpy/mass  is
\begin{equation} \label{4dNCM}
H \equiv M=\frac{\sqrt{\pi}}{4}\frac{r_+}{\gamma(r_+)}\left( 1+ \frac{r_+^2}{l^2} \right)  \,,
\end{equation}
while the black hole temperature is given by the known Hawking formula
\begin{equation} \label{4d_temp}
T=\frac{1}{4\pi}\frac{\mathrm{d} f(r)}{\mathrm{d} r} \Big|_{r=r_+}=\frac{1}{4\pi r_+} \left( 1+\frac{3r_+^2}{l^2} - r_+ \frac{\gamma'(r_+)}{\gamma(r_+)} -\frac{r_+^3}{l^2} \frac{\gamma'(r_+)}{\gamma(r_+)} \right) \,.
\end{equation}
In Fig.~\ref{fig:4dNCT}, we plot the temperature \eqref{4d_temp}.
For different values of $l\,$, the temperature appears two extrema (red solid curve in Fig.~\ref{fig:4dNCT}), or one inflexion point (black dashed curve in Fig.~\ref{fig:4dNCT}) where the local minimum coincides with the local maximum, or no extrema (blue solid curve in Fig.~\ref{fig:4dNCT}) with the temperature being  monotonically increasing. In contrast to the standard Schwarzschild-AdS black hole, each case considered here admits a zero-temperature extremal black hole remnant near $3 \sqrt{\theta}\,$,  rather than exhibiting a divergent behaviour at the origin. This ultraviolet completeness arises from additional quantum-corrected terms present in \eqref{4d_temp}. It is evident that noncommutativity regularizes the central singularity in much the same way that quantum mechanics resolves the Rayleigh-Jeans  catastrophe in black-body radiation.

\begin{figure}[h!] 
\includegraphics[width=0.53 \textwidth]{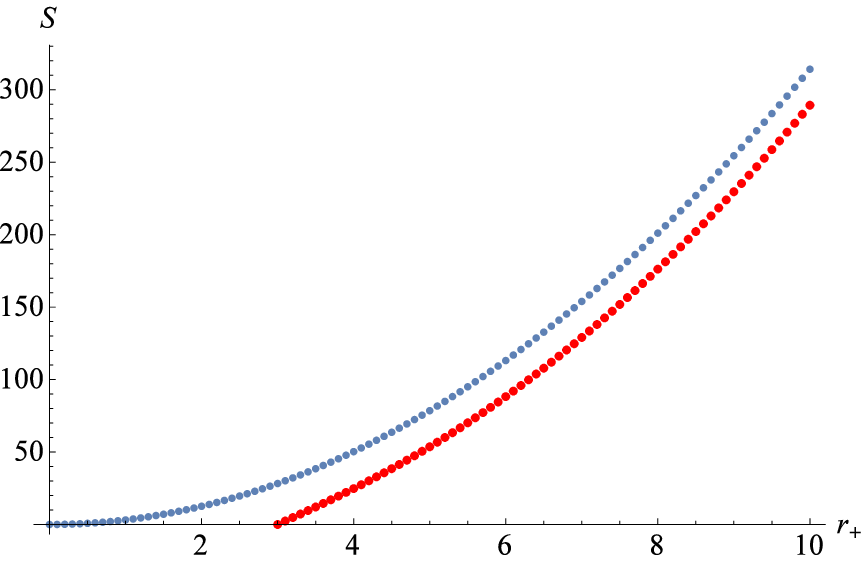}
\caption{The entropy is plotted for $\theta=1\,$. The red dots represent the noncommutative black hole entropy, while the blue dots represent the usual area law.}
\label{fig:4dNCS}
\end{figure}

The entropy of the black hole can be derived using thermodynamic arguments from the relation
\begin{equation} \label{thS}
  \mathrm{d}S=T \mathrm{d}M \,,
\end{equation}
while keeping  $P$ and $P_{\theta}$ constant, which corresponds to keeping  $l$ and $\theta$ constant, respectively. Integrating from the extremal black hole radius $r_0$ up to the event horizon $r_+\,$, we get the expression of the entropy
\begin{equation} \label{4dNCS}
S= \int \limits _{r_0}^{r_+} \mathrm{d} r \ \frac{\pi^{3/2} r}{\gamma(r)} \,.
\end{equation}
By writing $\gamma(r)=\frac{\sqrt{\pi}}{2}-\Gamma(r)\,$, we can Taylor expand \eqref{4dNCS}, approximating the entropy as
\begin{eqnarray} \nonumber
S &=& \int \limits _{r_0}^{r_+} \mathrm{d} r \ \frac{\pi^{3/2} r}{\frac{\sqrt{\pi}}{2} (1-\frac{2}{\sqrt{\pi}}\Gamma(r))} \\ \nonumber
&\simeq & \int \limits _{r_0}^{r_+} \mathrm{d} r \ 2\pi r \left(  1 + \frac{2}{\sqrt{\pi}} \Gamma(r)+ \mathcal{O}(\Gamma^2)\right) \\
& \simeq & \pi r^2 \left[ 1 - \frac{r}{\sqrt{\pi} \sqrt{\theta}} e^{-r^2/4\theta} + \frac{2 \Gamma(r)}{\sqrt{\pi}} \left( 1-\frac{6\theta}{r^2}\right)  \right] _{r_0}^{r_+} \label{ent_fl} \,
\end{eqnarray}
where quantum corrections to the area law become noticeable.  In Fig.~\ref{fig:4dNCS},  a numerical plot of the noncommutative  and the classical entropy is illustrated.
For large black holes ($r_+ \gg \sqrt{\theta}$), where $\gamma(r_+) \approx \sqrt{\pi}/2$ and $\gamma'(r_+) \approx 0$, the entropy coincides with the area law ($S \approx  \pi r_+^2$).

\begin{figure}[h!] 
\includegraphics[width=0.53 \textwidth]{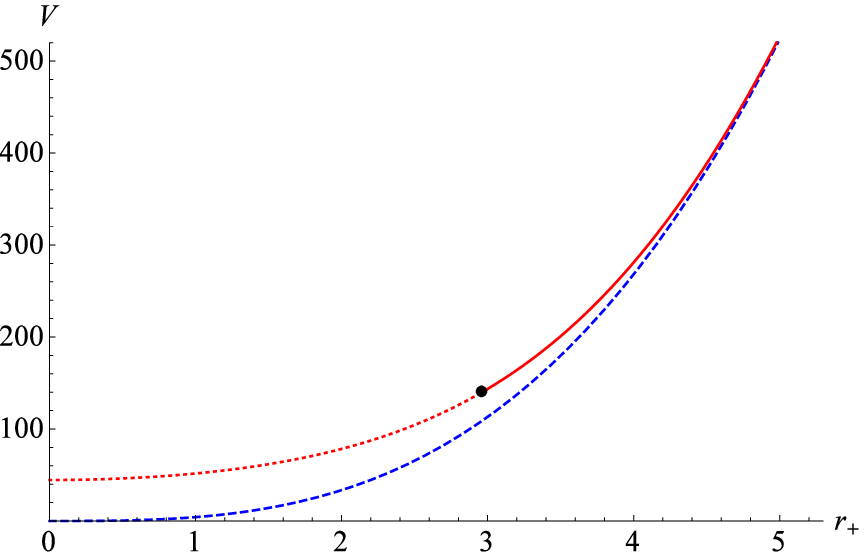}
\caption{The black hole volume   is displayed for $\theta=1\,.$ The red solid curve represents the volume of the noncommutative black hole, while the blue dashed curve represents the conventional volume $V_{\mathrm{AdS}} $ of a singular  black hole. The black dot spots the volume and the radius of the extremal black hole.}
\label{fig:4dNCV}
\end{figure}

\begin{figure}[h!] 
\includegraphics[width=0.53 \textwidth]{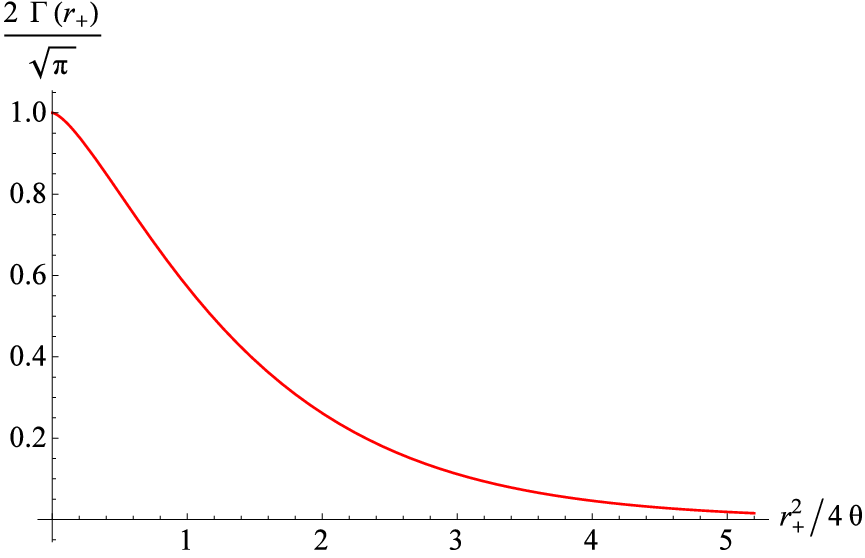}
\caption{The function $\frac{2}{\sqrt{\pi}}\Gamma(r_+)\,$ \textit{vs} $r_+^2/4\theta\,$.}
\label{fig:gamma}
\end{figure}

Taking into consideration the positive pressure of the system, which is given by
\begin{equation} \label{4d_pres}
P=\frac{3}{8\pi l^2} \,,
\end{equation}
we can  define the  black hole thermodynamic volume through the left equation  \eqref{volume} as
\begin{equation} \label{4dNCV}
V =\left(  \frac{\partial M}{\partial P}\right)_{P_{\theta},S}= \left(  \frac{\partial M}{\partial l}\right)_{\theta,r_+} \left(  \frac{\partial P}{\partial l}\right)^{-1}_{\theta,r_+} = \frac{4\pi r_+^3 }{3} \left(  \frac{\Gamma(3/2)}{\gamma(r_+)} \right) = V_{\mathrm{AdS}} \left(  \frac{\Gamma(3/2)}{\gamma(r_+)} \right)  \,
\end{equation}
where $\Gamma(3/2)=\sqrt{\pi}/2\,$,  $V_{\mathrm{AdS}}=\frac{4\pi r_+^3}{3} $ and the mass $M$ is given by \eqref{4dNCM}. Keeping $S$ constant implies that $r_+$ remains constant. We plot $V$ in Fig.~\ref{fig:4dNCV}.
Writing  $\gamma(r_+)$ in terms of the upper incomplete gamma function $\Gamma(r_+)=\Gamma(3/2,r_+^2/4\theta)\,$ as 
\begin{equation}
\gamma(r_+)=\frac{\sqrt{\pi}}{2} - \Gamma(r_+) \,,
\end{equation}
will help us see the correction terms in the  volume \eqref{4dNCV} relative to  $V_{\mathrm{AdS}}$ at short scales:
\begin{equation} \label{NCV}
V = \frac{2\pi^{3/2} }{3 \frac{\sqrt{\pi}}{2}\left(  1 - \frac{2}{\sqrt{\pi}} \Gamma(r_+)  \right) } r_+^3 \simeq \frac{4\pi r_+^3}{3} \left(  1 + \frac{2}{\sqrt{\pi}} \Gamma(r_+)+\frac{4}{\pi}\Gamma^2(r_+)+ \mathcal{O}(\Gamma^3) \right) \,.
\end{equation}
The above Taylor expansion is valid because $\Big| \frac{2}{\sqrt{\pi}} \Gamma(r_+)  \Big| < 1\,$, as can be verified graphically in Fig.~\ref{fig:gamma}.
The first-order approximation of \eqref{NCV} gives the conventional AdS volume $V_{\mathrm{AdS}} = \frac{4\pi}{3} r_+^3\,$, while the other terms represent thermodynamic corrections to $V_{\mathrm{AdS}}$ that arise due to quantum fluctuations of spacetime. We can estimate the black hole radius at which the noncommutative volume deviates significantly from the standard expression by numerically solving the equation $V = V_{\mathrm{AdS}}$. In other words, we need to solve the equation $\gamma(r_+) = \sqrt{\pi}/2\,$, which gives a numerical value of $r_+ \simeq 12.2 \sqrt{\theta}$. Thus, for $r_+ \gtrsim 12.2\sqrt{\theta}\,$, the noncommutative volume matches the usual AdS volume, while for $r_0 < r_+ < 12.2\sqrt{\theta}$, deviations appear. It is important to note that the volume is meaningless below the extremal black hole, as no black hole exists for $r_+ < r_0\,$.

Next, the noncommutative pressure is given by
\begin{equation} \label{4ncpress}
P_{\theta} = - \frac{3}{8\pi \theta} \,
\end{equation}
and the noncommutative volume by
\begin{equation} \label{4ncvol}
V_{\theta} = \left( \frac{\partial M}{\partial P_{\theta}} \right)_{P,S} =  \left( \frac{\partial M}{\partial \theta} \right)_{l,r_+} \left( \frac{\partial P_{\theta} }{\partial \theta} \right)^{-1}_{l,r_+}  = \frac{\pi^{3/2} r_+^4 (l^2+r_+^2) e^{-r_+^2/4\theta}}{12l^2 \sqrt{\theta} \ \gamma(3/2,r_+^2/4\theta)} \,.
\end{equation}
As we see, an important feature of the noncommutative black hole thermodynamics lies in the appearance of two distinct volume-pressure pairs, each contributing differently to the system's enthalpy. The standard thermodynamic volume is positive, reflecting the usual interpretation of space being carved out to accommodate the black hole, at an energy cost determined by the $PV$-term. 
The quantity $V_{\theta}$ indeed has dimensions of volume, implying a more subtle mechanism: the formation of the black hole is associated with a compressive effect driven by a self-gravitating, anisotropic matter distribution. This effect acts to reduce the effective volume of the system and can be understood as performing work on the black hole during its formation.  Notably, while the thermodynamic volume depends only on the noncommutative scale $\theta$, the noncommutative volume also involves the AdS curvature scale $l\,$. Furthermore, in the limit $\theta \rightarrow 0\,$, the noncommutative contributions vanish smoothly, ensuring consistency with the classical case.   On the other hand, the noncommutative pressure \eqref{4ncpress} must be negative, as it represents the self-gravitating droplet of anisotropic fluid performing work on the thermodynamic system by counteracting gravitational collapse (tension of the system). Consequently, the two pressures, $P$ and $P_{\theta}$, exert opposite effects on the system and thus have opposite signs.   This dual structure where two distinct pressure/volume pairs govern complementary thermodynamic aspects of the black hole, bears a conceptual resemblance to competing work terms in systems like evaporating droplets, where equilibrium is maintained by a balance between opposing forces.

\begin{figure}[h!] 
\includegraphics[width=0.55 \textwidth]{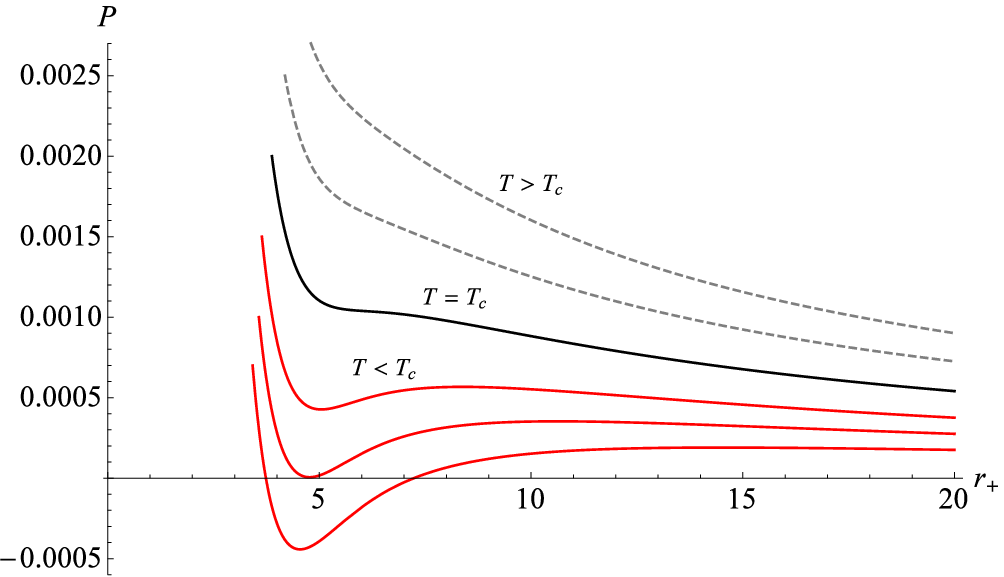}
\caption{The isotherms  of a ($3+1$)-dimensional noncommutative black hole for $\theta=1\,$. The red solid curves stand for $T<T_c$, the black solid curve for $T=T_c$ and the gray dashed curves for $T>T_c\,$.}
\label{fig:4dNCeos}
\end{figure}

Next, we  derive the equation of state  by combining \eqref{4d_temp} and \eqref{4d_pres}:
\begin{equation} \label{4dNCeos}
P=\frac{1}{1-\frac{r_+}{3} \frac{\gamma'(r_+)}{\gamma(r_+)}} \left( \frac{T}{2r_+} - \frac{1}{8\pi r_+^2} + \frac{\gamma'(r_+)}{8\pi r_+ \gamma(r_+)} \right) .
\end{equation}
In Fig.~\ref{fig:4dNCeos}, we plot the isotherms of the black hole and observe that, below a critical value of the temperature ($T<T_c$), the system exhibits an unstable non-physical branch that resembles a phase transition region  (red solid curves in Fig.~\ref{fig:4dNCeos}).
Thus, eq.~\eqref{4dNCeos} produces a pressure-volume diagram similar to that of a Van der Waals gas, provided one identifies the black hole variables $P,r_+$ and $T$ with the pressure, the specific volume and the temperature of the  gas  respectively. The extended phase space improves the  analogy between the  black hole thermodynamic variables  with the Van der Waals gas variables, relative to the non-extended phase space  where $P$ and $1/T$ of the  black hole are instead identified with the temperature and the pressure of the  gas \cite{NiT11}.

\begin{figure}[h!] 
\includegraphics[width=0.53 \textwidth]{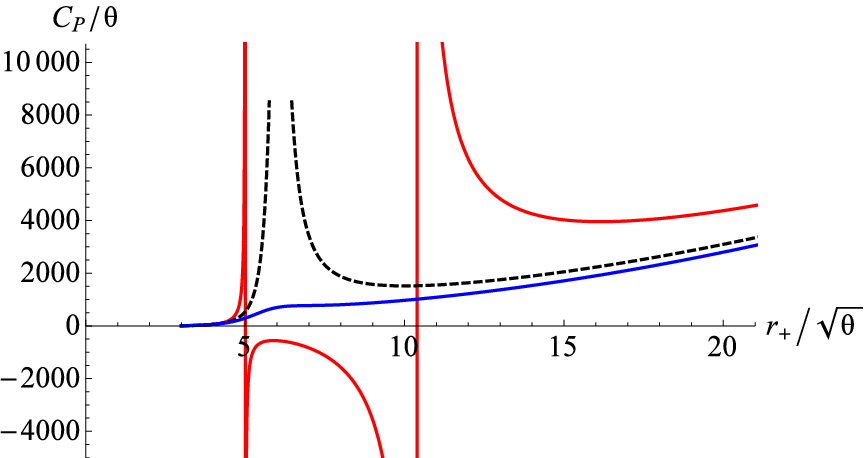}
\caption{The specific heat   of a ($3+1$)-dimensional noncommutative black hole  is displayed for $l=18 \sqrt{\theta}$ (red solid curve), for $l=l_c\approx 11.15 \sqrt{\theta}$ (black dashed curve) and for $l=8 \sqrt{\theta}\,$ (blue solid curve).}
\label{fig:4dNCCp}
\end{figure}

Finally, we calculate the two specific heats and the Gibbs energy. Following the definitions \eqref{specific_heat}, we retrieve $C_V=0$ and
\begin{equation}
C_P =\frac{\pi^{3/2} r_+^2 \left[ \gamma'(r_+) (l^2+r_+^2)r_+ - \gamma(r_+) (l^2+3r_+^2) \right] }{(l^2-3r_+^2) \gamma^2(r_+) - r_+^2 (l^2+r_+^2) \gamma'^2(r_+)+ r_+^2 \gamma(r_+) \left[ 2r_+ \gamma'(r_+) + (l^2+r_+^2) \gamma''(r_+) \right] } \,.
\end{equation}
In Fig.~\ref{fig:4dNCCp}, we plot $C_P$ for different values of $l$ and we see the divergent regions where the  phase transition  takes place.
For values of $l$ greater than the critical one $(l>l_c \approx 11.15 \sqrt{\theta})$, we have two divergent regions for $C_P$   and three distinct branches (red solid curve in Fig.~\ref{fig:4dNCCp}). The first branch  corresponds to a small stable black hole with $C_P>0\,$, the middle branch to an unstable black hole with $C_P<0$ and the last one  to a large stable black hole with $C_P>0\,$. In other words, a first-order phase transition seems to occur between a small/large black hole reminiscent of the liquid/gas coexistence of a Van der Waals gas. For $l=l_c\,$, there exists only one divergence in $C_P\,$,  where the radius of the small and the large black hole coincide at one inflexion point (black dashed curve in Fig.~\ref{fig:4dNCCp}). For $l<l_c\,$, no phase transitions occur and there exists a single thermally stable black hole (blue solid curve in Fig.~\ref{fig:4dNCCp}). Noncommutativity modifies the heat capacity profile, giving rise to three distinct cases, in contrast to the classical AdS counterpart, which exhibits only a single phase transition between background radiation and a large stable black hole, known also as \textit{Hawking-Page transition}.

\begin{figure}[h!] 
\includegraphics[width=0.53 \textwidth]{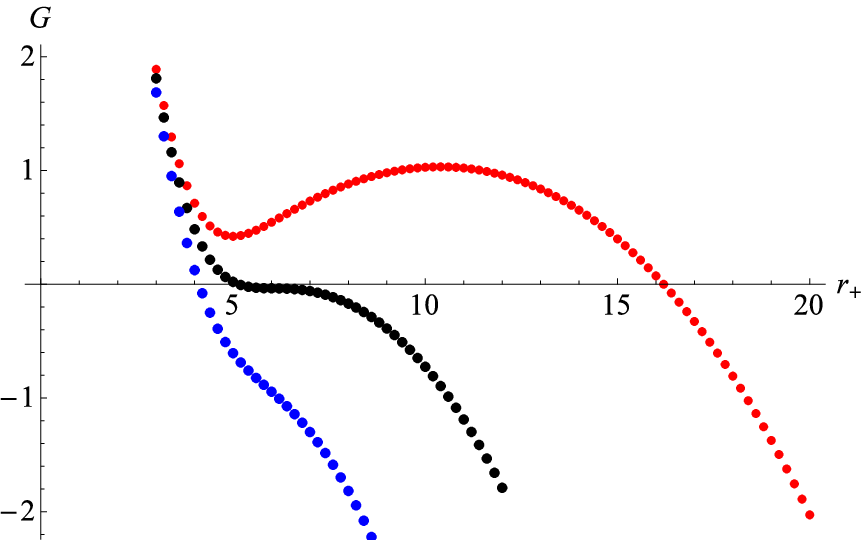}
\caption{The Gibbs energy  of a ($3+1$)-dimensional noncommutative black hole  for $l=18 \sqrt{\theta}$ (red dots), for $l=l_c \approx 11.15 \sqrt{\theta}$ (black dots) and for $l=8\sqrt{\theta}$ (blue dots). The plot  is displayed for $\theta=1\,.$}
\label{fig:4dNCG}
\end{figure}

Proceeding with the Gibbs energy, we insert \eqref{4dNCM}, \eqref{4d_temp} and \eqref{4dNCS} into $G=M-T S\,$. We plot numerically in Fig.~\ref{fig:4dNCG} the form of $G\,$.
For $l>l_c\,$, there are two extrema for $G$ (red dots in Fig.~\ref{fig:4dNCG}), one minimum and one maximum, representing the radii of the two stable black holes (small/large). For $l=l_c\,$, the two extrema merge at one inflexion point (black dots in Fig.~\ref{fig:4dNCG}), while for $l<l_c\,$, there are no phase transitions (blue dots in Fig.~\ref{fig:4dNCG}) and the preferred states are those with the smaller Gibbs energy, i.e., with the larger black hole radius. A comparison with the classical AdS black hole is instructive here. In the classical case, the Gibbs free energy exhibits a single maximum before decreasing toward more thermodynamically stable configurations, reflecting the sole presence of the Hawking-Page transition. In contrast, noncommutative geometry leads to a richer phenomenology, allowing for the existence of additional globally stable remnants beyond the standard phase structure.

\section{Noncommutative black hole chemistry in ($2+1$)-dimensions} 
\label{3dNC_chem}

\subsection{The line element} 
\begin{figure}[h!] 
\includegraphics[width=0.53 \textwidth]{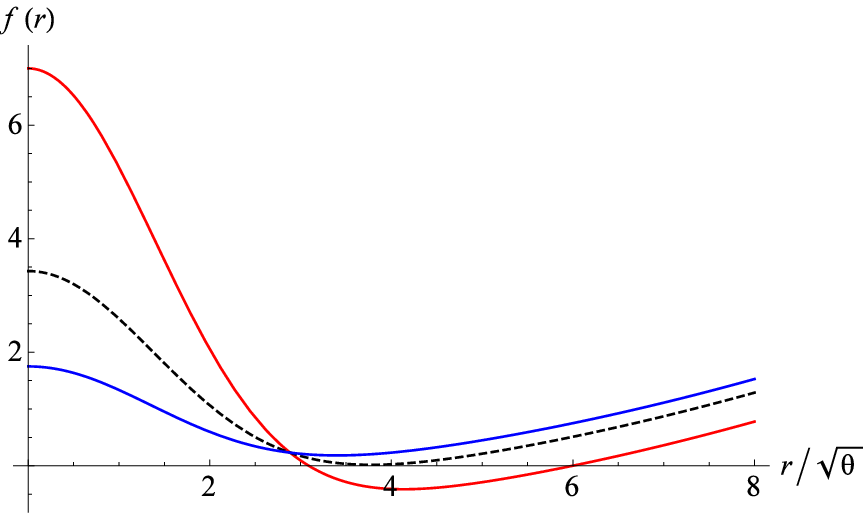}
\caption{The NCBTZ metric potential $f(r)$ is displayed for fixed $l=6\sqrt{\theta}$ and for $M=1.1$ (red solid curve), for $M=M_0=0.51$ (black dashed curve) and for $M=0.25$ (blue solid curve). }
\label{fig:3dNCMetric}
\end{figure}

We are interested in finding the line element of a circular symmetric metric of the form
\begin{equation} \label{3d_NC_line_element}
\mathrm{d} s^2 = -f(r) \mathrm{d} t^2 + f(r)^{-1} \mathrm{d} r^2 + r^2 \mathrm{d} \phi^2 \,
\end{equation}
under the assumption of noncommutativity of spacetime \cite{RKB++13}. For $d=2$ in \eqref{NVdensity}, the profile of the density reads
\begin{equation} \label{3D_density}
\rho(r) = \frac{M}{4\pi \theta} \exp({-r^2/4\theta}) \,.
\end{equation}
From the vacuum condition $g_{tt} = -1/ g_{rr}$ and from the covariant conservation $\nabla_{\mu} \mathcal{T}^{\mu}_{\nu}=0\,$, we specify the components of the stress-energy tensor:
\begin{equation} \label{3d_NC_stress_tensor}
\mathcal{T}^{\mu}_{\nu}=\mathrm{diag} \left( -\rho(r), \ p_{\mathrm{r}}(r), \ p_{\perp}(r) \right) \,
\end{equation}
with
\begin{equation}
p_{\mathrm{r}}(r)=-\rho(r) \quad \mathrm{and} \quad  p_{\perp}(r)= - \rho(r) - r \frac{\partial \rho(r)}{\partial r}  \,.
\end{equation}
By solving the Einstein field equations \eqref{einstein_eq} with the above line element \eqref{3d_NC_line_element} and the   tensor \eqref{3d_NC_stress_tensor}, we get the metric potential 
\begin{equation} \label{3dsolution}
f(r) = C + 8M G_{(3)} e^{-r^2/4\theta} - \Lambda r^2 \,
\end{equation}
 with $G_{(3)}$ being the ($2+1$)-dimensional Newton's constant  and $C$ an integration constant. We require that the solution \eqref{3dsolution} matches the conventional BTZ black hole \cite{BTZ92,Car95} at large scales, where quantum fluctuations are negligible. In other words, at $r/\sqrt{\theta} \rightarrow \infty\,$, we have $e^{-r^2/4\theta} \rightarrow 0$ and thus $f(r) \rightarrow -M G_{(3)}-\Lambda r^2\,$. Therefore, we can identify  $C=-M G_{(3)}\,$ and, in units where $c=\hbar=k_{\rm B}=G_{(3)}=1\,,$ the metric potential of a NCBTZ black hole  reads
\begin{equation} \label{NCBTZ}
f(r) = -M + 8M e^{-r^2/4\theta} + \frac{r^2}{l^2} \,.
\end{equation}
 The positive AdS radius $l$ in $(2+1)$-dimensions is related to $\Lambda$ through the relation $\Lambda=-1/l^2\,$. The potential \eqref{NCBTZ} has three different horizon structures depending on the values of $M$ and $l\,,$ as can be seen in Fig.~\ref{fig:3dNCMetric}:
\begin{itemize}
\item Two horizons for $M>M_0$ (red solid curve in Fig.~\ref{fig:3dNCMetric}); one Cauchy  $r_-$ and one event   horizon $r_+\,$.
\item One degenerate horizon (black dashed curve in Fig.~\ref{fig:3dNCMetric})  representing the extremal black hole with the lowest possible mass $M=M_0$ and radius $r_0=r_-=r_+\,$.
\item No  horizons for $M<M_0$ (blue solid curve in Fig.~\ref{fig:3dNCMetric}).
\end{itemize}
In the opposite limit where $r \ll \sqrt{\theta}$ (near the origin), we can approximate $e^{-r^2/4\theta} \approx 1 - \frac{r^2}{4\theta}\,$, allowing the metric \eqref{NCBTZ} to take the approximate form of
\begin{equation} \label{3dorigin}
f(r) \approx 7M - \Lambda_{\mathrm{eff}} r^2 \,,
\end{equation}
where $\Lambda_{\mathrm{eff}} = \frac{2M}{\theta} - \frac{1}{l^2}\,$. Thus we can distinguish 3 cases:
\begin{itemize}
\item $\Lambda_{\mathrm{eff}} > 0 $ if $M > \frac{\theta}{2l^2}\,.$
\item $\Lambda_{\mathrm{eff}} < 0 $ if $M < \frac{\theta}{2l^2}\,.$
\item $\Lambda_{\mathrm{eff}} = 0 $ if $M = \frac{\theta}{2l^2}\,.$
\end{itemize}
The condition $M \geq M_0 \geq \frac{\theta}{2l^2}$ must hold in \eqref{3dorigin},  in order for horizons to exist. This implies $\Lambda_{\rm{eff}} \geq 0$  near the origin.

\subsection{The chemistry}

\begin{figure}[h!] 
\includegraphics[width=0.53 \textwidth]{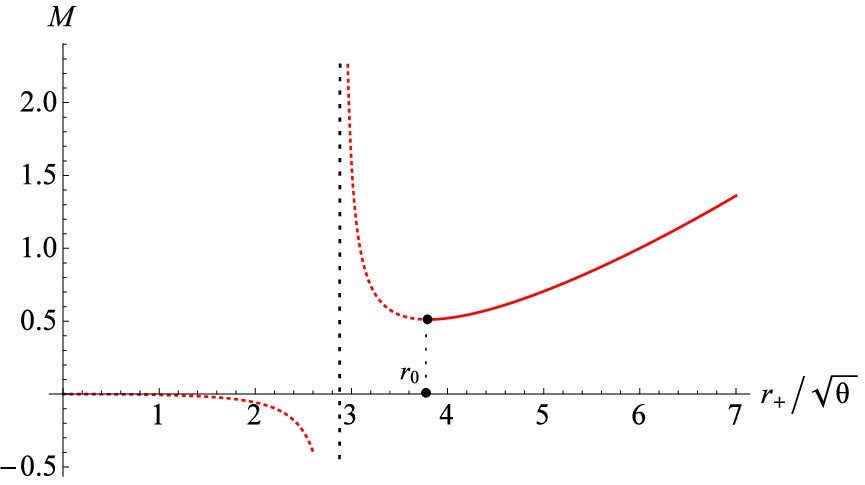}
\caption{The mass  of the NCBTZ black hole  is displayed for $l=6\sqrt{\theta}\,$ and $r_+>r_0 \simeq 3.789 \sqrt{\theta}\,$.}
\label{fig:3dNCM}
\end{figure}

We now study  thermodynamic properties of the metric \eqref{NCBTZ} within the extended phase space, with the pressure terms given by \eqref{press} for $d=2$:
\begin{equation} \label{3dNCP}
P=\frac{1}{8\pi l^2} \qquad \mathrm{and} \qquad  P=-\frac{1}{8\pi \theta} \,.
\end{equation}
From the relation $f(r_+)=0\,$,  we express the black hole mass as
\begin{equation}
M=\frac{r_+^2}{l^2(1-8 e^{-r^2/4\theta})} \,.
\end{equation}
We imply a positive  mass and so the relation $1-8 e^{-r^2/4\theta} > 0$ must hold, giving the constraint $r_+ > 2.88 \sqrt{\theta}$ for the black hole radius. This can be seen graphically in Fig.~\ref{fig:3dNCM}.
From the relation $\frac{\mathrm{d} M}{\mathrm{d} r_+}=0$\,, we find  that the lowest black hole mass for the specific example in Fig.~\ref{fig:3dNCM} is $M_0 \simeq 0.51\,$, having a  degenerate horizon  at $r_0 \simeq 3.789\sqrt{\theta}\,$. Therefore,  a black hole exists for the range $r_+ \gtrsim 3.789\sqrt{\theta}\,$.

\begin{figure}[h!] 
\includegraphics[width=0.53 \textwidth]{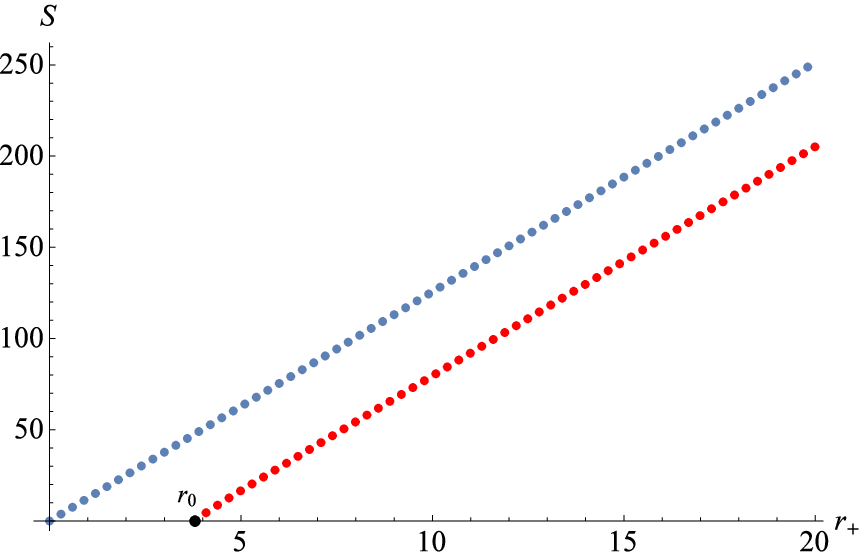}
\caption{The entropy plot is displayed for $\theta=1\,$. The red dots represent the NCBTZ black hole entropy, while the blue dots the usual BTZ black hole entropy $S_{\mathrm{BTZ}}\,$.}
\label{fig:3dNCS}
\end{figure}

The black hole entropy can be derived from \eqref{1stLaw} and is given by
\begin{equation} \label{3dNCS}
S = \int \limits_{r_0}^{r_+} \mathrm{d} r \ \frac{4\pi}{1-8 e^{-r_+^2/4\theta}} \,.
\end{equation}
In the classical limit ($\frac{r_+}{\sqrt{\theta}} \rightarrow \infty$),  the smallest radius becomes negligible ($r_0 \rightarrow 0$) and  the entropy \eqref{3dNCS} matches the usual BTZ black hole entropy which is given by twice the circumference $L$ of the black hole \cite{BTZ92}, i.e.,  $S \approx S_{\mathrm{BTZ}}=4\pi r_+=2L\,$. A numerical plot of the  entropy is illustrated in Fig.~\ref{fig:3dNCS}. In the noncommutative case (depicted by red dots), the plot highlights the presence of a zero-entropy black hole remnant at $r_0\,$, which contrasts with the classical scenario where no such remnant appears.

\begin{figure}[h!] 
\includegraphics[width=0.53 \textwidth]{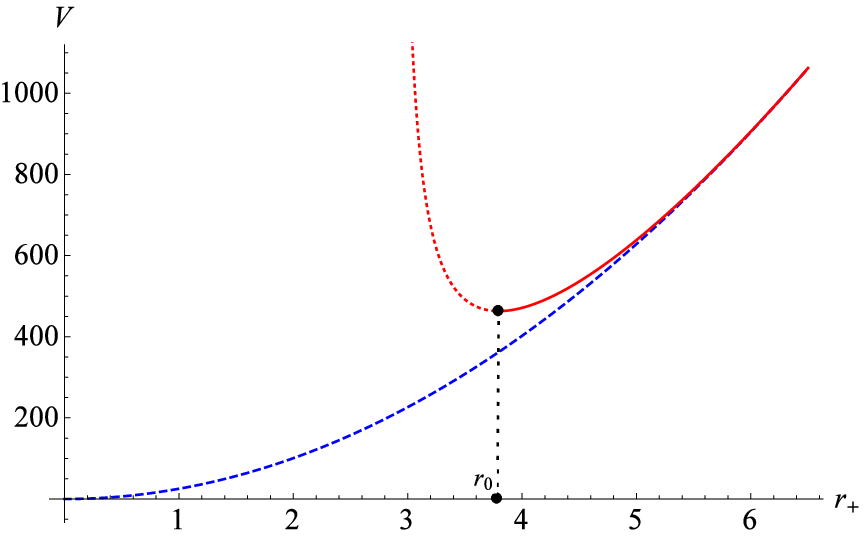}
\caption{The red curve represents the NCBTZ black hole volume for $r_+ > r_0 \simeq  3.789 \sqrt{\theta}\,$, while the blue dashed curve represents the classical volume $V_{\mathrm{BTZ}}$. The plot is displayed for $\theta=1\,.$}
\label{fig:3dNCV}
\end{figure}

\begin{figure}[h!] 
\includegraphics[width=0.53 \textwidth]{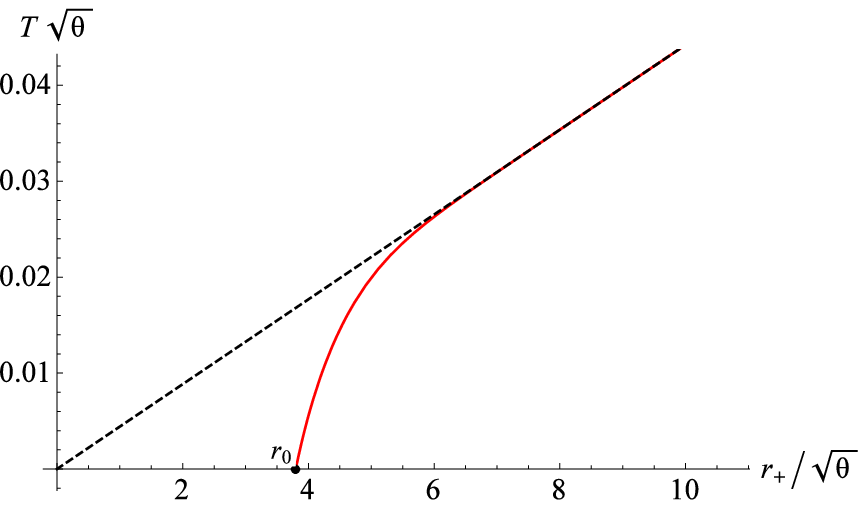}
\caption{The red curve represents the temperature  of the NCBTZ black hole    for fixed $l=6\sqrt{\theta}$ and $r_+>r_0 \simeq 3.789 \sqrt{\theta}\,$. The black dashed curve shows $T_{\mathrm{BTZ}}\,.$}
\label{fig:3dNCT}
\end{figure}

Next, from \eqref{volume} we can define the black hole  ``volume", which is actually a surface in ($2+1$)-dimensions, as
\begin{equation}
V=  \frac{8\pi r_+^2}{1-8e^{-r_+^2/4\theta}}  = \frac{V_{\mathrm{BTZ}}}{1-8e^{-r_+^2/4\theta}} \,
\end{equation}
where $V_{\mathrm{BTZ}}=8\pi r_+^2$ is the volume of the classical  BTZ black hole. Its shape is plotted in Fig.~\ref{fig:3dNCV}. 
The noncommutative volume reads
\begin{equation}
V_{\theta} =  \frac{16\pi r_+^4 e^{-r_+^2/4\theta}}{l^2 (1 - 8e^{-r_+^2/4\theta})^2} 
\end{equation}
and is always positive. By using the Hawking formula for the temperature, we find 
\begin{equation} \label{3dNCT}
T= \frac{r_+}{2\pi l^2} \left(  1 - \frac{2r_+^2/\theta}{ e^{r_+^2/4\theta}-8 } \right) .
\end{equation}
The first term of the temperature \eqref{3dNCT} corresponds to the standard  temperature of the BTZ black hole, whereas the second term represents the noncommutative thermodynamic correction. This correction vanishes at large distances, leading to the approximation $T \approx T_{\mathrm{BTZ}}=\frac{r_+}{2\pi l^2}\,$. The form of the temperature is plotted in Fig.~\ref{fig:3dNCT}. The temperature is positive for $r \gtrsim 3.789\sqrt{\theta}$ and monotonically increasing with the increase of the horizon.

\begin{figure}[h!] 
\includegraphics[width=0.53 \textwidth]{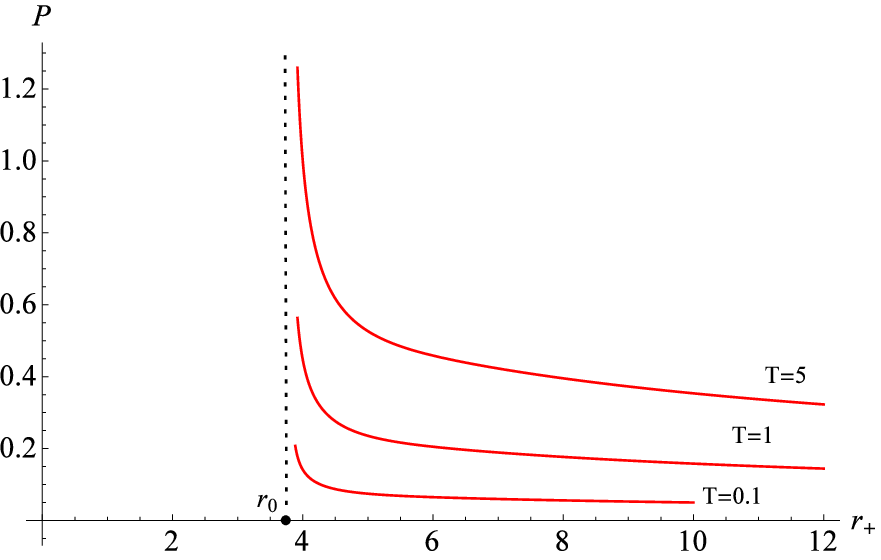}
\caption{The isotherms of the NCBTZ black hole are displayed for $\theta=1\,$.}
\label{fig:3dNCeos}
\end{figure}

Next, we obtain the equation of state of the NCBTZ black hole by combining \eqref{3dNCT} with \eqref{3dNCP}:
\begin{equation} \label{3dNCeos}
P =\sqrt{ \frac{T}{4r_+ \left( 1 - \frac{2r_+^2/\theta}{e^{r_+^2/4\theta}-8} \right) } }\,.
\end{equation}
Eq.~\eqref{3dNCeos} admits no critical behaviour and the isotherms of the black hole are illustrated in Fig.~\ref{fig:3dNCeos}.

\begin{figure}[!h] %
\centering
\subfigure[The specific heat $C_P\,$. ]{%
\label{fig:3dNCCp}%
\fbox{\includegraphics[height=1.9in]{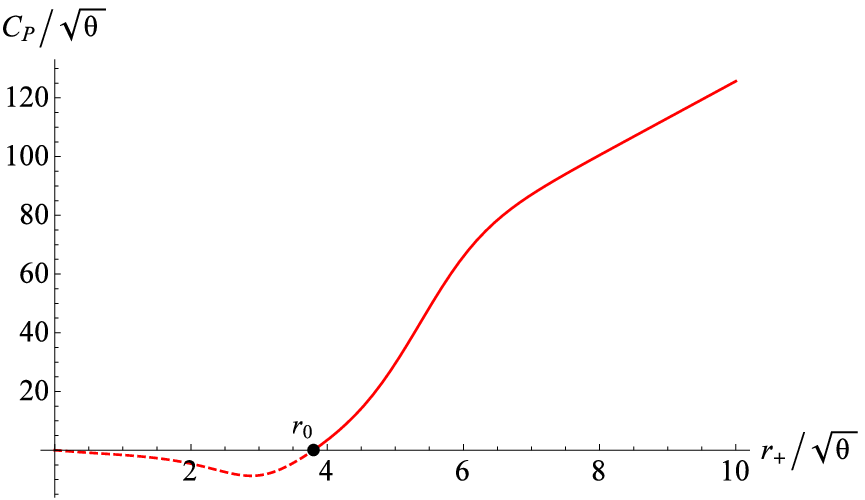}}}%
\qquad
\subfigure[The Gibbs energy $G$  for $\theta=1\,.$]{%
\label{fig:3dNCG}%
\fbox{\includegraphics[height=1.9in]{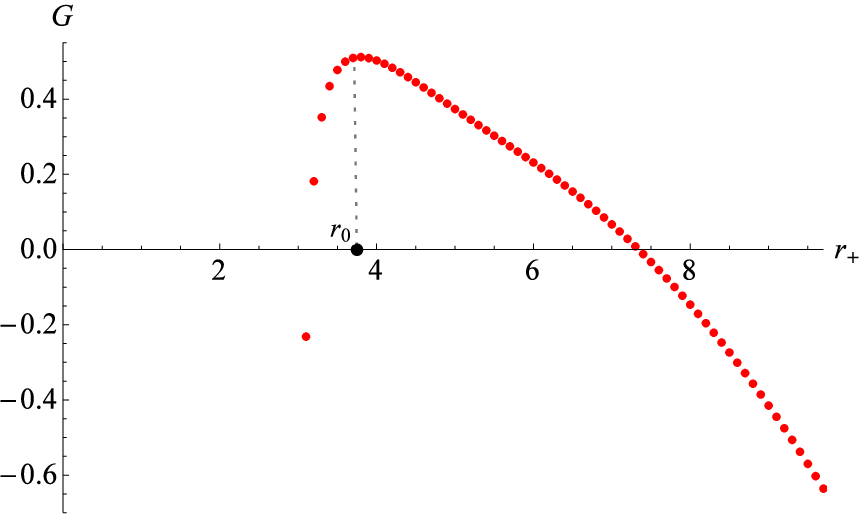}}}%
\caption{The plots of $C_P$ and $G$ for the NCBTZ black hole are displayed  for fixed $l=6 \sqrt{\theta} \,$.}
\label{fig:3dNCCpG}
\end{figure}

Lastly, we evaluate the two specific heats and the Gibbs energy. The two specific heats are
\begin{equation}
C_{V} = 0 
\quad \mathrm{and} \quad C_P = \frac{4\pi \theta r_+ e^{r_+^2/4\theta}  \left[ \left( e^{r_+^2/4\theta}-8 \right)\theta - 2r_+^2  \right] }{ r_+^4 e^{r_+^2/4\theta} - 6\theta r_+^2 \left( e^{r_+^2/4\theta}-8\right) + \left( e^{r_+^2/4\theta}-8\right)^2 } \,, 
\end{equation}
while the Gibbs energy is given by the relation $G=M-T S \,$. The plots of $C_P$ and $G$ are displayed in Fig.~\ref{fig:3dNCCpG}. For values $r_+>r_0\,$, the specific heat is positive and monotonically increasing, while the Gibbs energy is monotonically decreasing with the increase of the black hole radius. This indicates that the NCBTZ black hole is thermodynamically  stable both locally and globally, similar to its singular counterpart\cite{FMM15}. There are no signs of phase transitions, and larger black holes are energetically favoured.

\section{Noncommutative black hole chemistry in ($1+1$)-dimensions} 
\label{2dNC_chem}

The specific dimensionality requires  special treatment due to the triviality of the Einstein tensor. Stating it differently, the Einstein tensor vanishes for every form of the metric tensor, making ($1+1$)-dimensional dynamics meaningless \cite{Col77}. For this reason, various attempts have been made over the years \cite{MST90,Sch99,GKV02,GrM07} to formulate ($1+1$)-dimensional gravity with the help of a dilaton field $\psi\,$, which is strongly coupled to gravity. It has also been shown \cite{MaR93} that the  limit, $D \rightarrow 2\,,$ of the D-dimensional Einstein-Hilbert action leads to the following dilatonic action 
\begin{equation} \label{a5}
\mathcal{S}=   \int \mathrm{d} ^2x \sqrt{-g}\ \left[ \frac{1}{16\pi G_{(2)}}\left(  \psi R + \frac{1}{2}  (\nabla \psi)^2 - 2\Lambda \right)+ \mathcal{L}_{\mathrm{m}} \right] \,
\end{equation} 
where $G_{(2)}$ is the ($1+1$)-dimensional Newton's constant.  The variation of the action \eqref{a5} with respect to the two dynamical fields ($\psi$ and $g_{\mu\nu}$) leads to the Liouville field equation
\begin{equation} \label{a9}
R + 2\Lambda= 8\pi G_{(2)} \mathcal{T} \,
\end{equation}
which is considered  to be so far the best analogue of Einstein field equations in ($1+1$)-dimensions. Eq.~\eqref{a9} offers a variety of rich characteristics, such as black holes \cite{Man94}, $2D$ FRW-like Cosmologies \cite{ChM95}, black hole radiation \cite{MaS92} and black hole nucleation \cite{TNM++18}. Before proceeding, we must highlight three ``peculiarities'' arising from the nature of this dimensionality: 
\begin{enumerate}
\item  The reduced Newton's constant $G_{(2)}$ is dimensionless and admits the value of $2\pi$ in natural units  \cite{MuN13}.  Hence, no Planck scale exists to distinguish the quantum regime from the classical one. We will use this value for $G_{(2)}$ in the following calculations.
\item In contrast to higher dimensions, mass scales inversely with distance ($M \sim x^{-1}$) meaning that a heavier black hole is smaller in size and vice versa. In other words, a black hole behaves like an elementary particle \cite{SpS16}. 
\item  There seems to be a sign-reversal in $\Lambda$ relative to the higher dimensions; when $\Lambda > 0\,$, it yields  AdS asymptotics,  while $\Lambda < 0$ characterizes a  de Sitter black hole. For further details about the sign-reversal of $\Lambda\,$, see \cite{FMM15,TNM++18}. 
\end{enumerate}
To understand these reversed behaviours, which stand in contrast to those observed in conventional higher-dimensional settings, one must begin with the application of Gauss's theorem in lower dimensions. More specifically,
\begin{equation}
\oint \vec{g}_{(2)} \mathrm{d}\vec{S}^{(0)} = 2 G_{(2)} M \quad \Rightarrow \quad g_{(2)}= G_{(2)} M \,
\end{equation}
where $g_{(2)}$ is the gravitational vector field, or, the gravitational acceleration, due to
the mass M, which is enclosed inside a zero-sphere $S^{(0)}$. The zero-sphere can be thought as a pair of two antipodal points, which are the boundary of a line segment. This geometric structure implies that, in $(1+1)$-dimensional spacetime, the Newtonian force experienced by a test particle is constant along the entire line. This behavior contrasts sharply with the conventional case, where the Newtonian force is inversely proportional to the square of the distance between two masses. As a consequence, the gravitational potential in these dimensions, obtained by integrating the constant force field, is linear in the distance $x$ from the source. Therefore the classical line element of a $(1+1)$-dimensional AdS black hole is found to be 
\begin{equation}
f(x) = C + 4\pi M|x| + \frac{x^2}{l^2} \,
\end{equation}
where $C$ is an integration constant introduced by the original authors \cite{MST90}, whose value characterizes different classes of manifolds.

\subsection{The line element} 

\begin{figure}[h!] 
\begin{center}
\includegraphics[width=0.53 \textwidth]{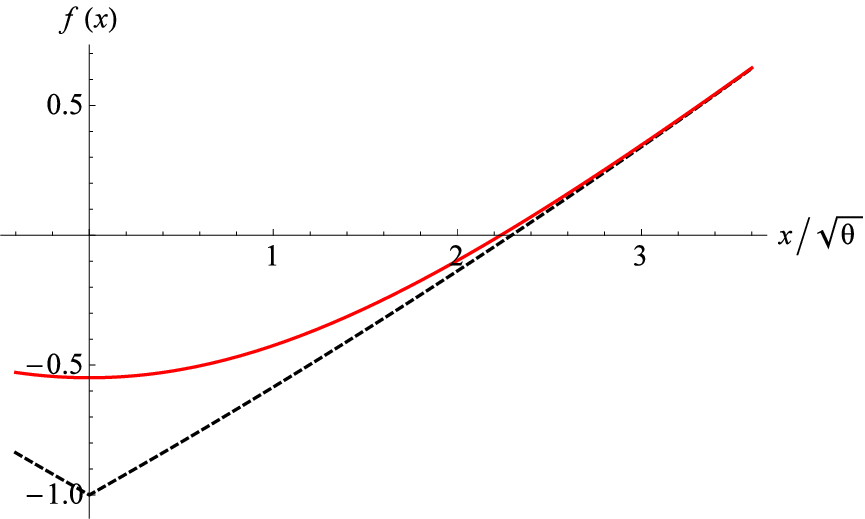}
\caption{The red solid curve represents the  metric potential \eqref{2dNCAdS_potential}  for $M\sqrt{\theta}=0.2 $ and $l=8 \sqrt{\theta}\,$. The black dashed curve represents the analogous singular black hole case. }
\label{fig:2dNCMetric}
\end{center}
\end{figure}

The  solution of a ($1+1$)-dimensional noncommutative black hole along with its causal structures have been extensively   discussed   in \cite{MuN11} where the authors present	 an extension of the work in \cite{MST90}.  Here, we provide a brief overview of the solution we are focusing on. The authors  of the corresponding paper followed the approach of the nonlocal deformation of the action \eqref{a5} by means of the operator $e^{\theta \square}$ \cite{MMN11}, where $\square=\nabla^2\,$. In this case, the Ricci scalar in the action is replaced  by
\begin{equation}
R \rightarrow \mathcal{R} = \int \mathrm{d} ^2y \sqrt{-g} \ \mathcal{A}(x-y) R(y) \,
\end{equation}
where
\begin{equation}
\mathcal{A}(x-y) = e^{-\theta \square} \delta^{(2)} (x-y) \,. 
\end{equation}
From the variation of the new  action, we conclude to the following noncommutative Liouville field equation:
\begin{equation} \label{NC_Liouville}
R+2\Lambda=8\pi G_{(2)} e^{ \theta \square} \mathcal{T} \,.
\end{equation}
The new profile for the energy density  will be
\begin{equation} \label{2dNCdensity}
\rho(x) = \left( \frac{M}{2\pi } \right)  \frac{1}{\sqrt{4\pi \theta}} \exp(-x^2/4\theta) \,
\end{equation}
where we have normalised \eqref{NVdensity} with a factor of $2\pi$ in order to be consistent with the work in \cite{MST90}. The trace $\mathcal{T}$  can be determined by imposing a perfect fluid with a stress-energy tensor of the form
\begin{equation}
\mathcal{T}^{\mu\nu} = (\rho(x) + p(x)) u^{\mu}u^{\nu} + p(x) g^{\mu\nu} \,,
\end{equation}
where $u^{\mu}$  is a normal timelike vector for a fluid at rest, $p(x)$ is the pressure term of the fluid and $\rho(x)$ is given by \eqref{2dNCdensity}. For a line element of the form
\begin{equation}
\mathrm{d} s^2 = - f(x) \mathrm{d} t^2 + \frac{\mathrm{d} x^2}{f(x)} \,,
\end{equation}
the Ricci scalar is $R=-\frac{\mathrm{d} ^2f(x)}{\mathrm{d} x^2}$ and so \eqref{NC_Liouville} becomes
\begin{equation} \label{NC_Liouville2}
\frac{\mathrm{d} ^2f(x)}{\mathrm{d} x^2} - 2\Lambda =  8\pi G_{(2)} (\rho(x) - p(x)) \,.
\end{equation}
Requiring a reflection symmetry around the origin ($x \rightarrow |x|$), analogous to spherical symmetry in higher dimensions, we can write \eqref{NC_Liouville2} as
\begin{equation} \label{NC_Liouville3}
f''(x) +2f'(x) \delta(x) - 2\Lambda =  8\pi G_{(2)} (\rho(x) - p(x)) \,
\end{equation}
where $f'(x) = \frac{\mathrm{d} f(x)}{\mathrm{d} |x|}\,$.
In order to solve \eqref{NC_Liouville3}, we need to use the equation of hydrostatic equilibrium resulting from the conservation of $\mathcal{T}^{\mu\nu}$  ($\nabla_{\mu} \mathcal{T}^{\mu\nu}=0$):
\begin{equation} \label{NC_HE}
\frac{\mathrm{d} p(x)}{\mathrm{d} x} = - \frac{1}{2} \left( \frac{\mathrm{d}}{\mathrm{d} x} \ln f(x) \right) (\rho(x)+p(x)) \,.
\end{equation}
We cannot solve the noncommutative Liouville equation in an exact analytic form, but we can make the approximation $p(x) \approx 0\,$, since noncommutative effects are relevant at short distances, and the region where $p(x)$ is non-zero is limited. Therefore, solving \eqref{NC_Liouville3} in conjunction with \eqref{NC_HE} for $p(x) \approx 0\,,$ we obtain the leading order solution 
\begin{equation}
f(x) = C + \Lambda x^2 + \frac{8\pi M \sqrt{\theta}}{\sqrt{\pi}} \left( \exp\left( -\frac{x^2}{4\theta}\right) + \frac{|x|}{2\sqrt{\theta}} \gamma \left( 1/2 , x^2/4\theta \right)    \right) \,
\end{equation}
where
\begin{equation}
\gamma(1/2 , x^2/4\theta) = \int \limits ^{x^2/4\theta}_{0} \mathrm{d} z \ z^{-1/2} e^{-z} \,.
\end{equation}
This solution provides  a  variety of spacetimes with multiple horizons \cite{MuN11}. However, to obtain the corresponding AdS black hole necessary for studying the chemistry,   we must set $C=-1$ and $\Lambda = 1/l^2\,$, where $l$ is the AdS radius. Then, the metric potential of a ($1+1$)-dimensional noncommutative anti-de Sitter  black hole will be
\begin{equation} \label{2dNCAdS_potential}
f(x) = -1 + \frac{x^2}{l^2} + \frac{8\pi M \sqrt{\theta}}{\sqrt{\pi}} \left( \exp\left( -\frac{x^2}{4\theta}\right) + \frac{|x|}{2\sqrt{\theta}} \gamma \left( 1/2 , x^2/4\theta \right)    \right) \,.
\end{equation}
 In the classical limit ($|x| \gg \sqrt{\theta}$), the gamma function is approximated by $\sqrt{\pi}$ and the potential \eqref{2dNCAdS_potential} coincides with the singular AdS black hole, i.e., $f(x) \approx -1 + 4\pi M|x| + \frac{x^2}{l^2}\,$.   
Near the origin ($|x| \ll \sqrt{\theta}$), the gamma function is approximately $ |x|/\sqrt{\theta}\,$ and so the metric potential becomes
\begin{equation}
f(x) \approx \frac{4M \sqrt{\theta}}{\sqrt{\pi}} - 1 + \left( \frac{M}{\sqrt{\pi \theta}} + \frac{1}{l^2} \right) x^2 \,.
\end{equation}
This means that noncommutativity has smoothed the local profile of the solution which now has a quadratic dependence on $|x|\,$, instead of the linear dependence observed in the classical case. As a result, the Ricci scalar remains finite everywhere. In Fig.~\ref{fig:2dNCMetric}, we plot $f(x)$ and observe that noncommutative geometry has smoothed the profile of $f(x)$ near the origin, relative to the singular case, thus providing a regular black hole.

\subsection{The chemistry} 

\begin{figure}[h!] 
\begin{center}
\includegraphics[width=0.53 \textwidth]{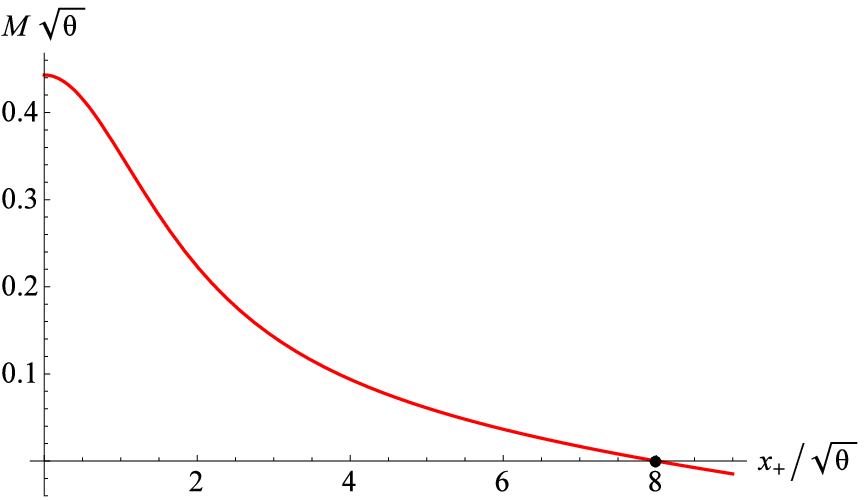}
\caption{The mass  of the $(1+1)$-dimensional noncommutative  black hole    for  $l=8\sqrt{\theta}\,$. }
\label{fig:2dNCM}
\end{center}
\end{figure}

\begin{figure}[h!] 
\begin{center}
\includegraphics[width=0.53 \textwidth]{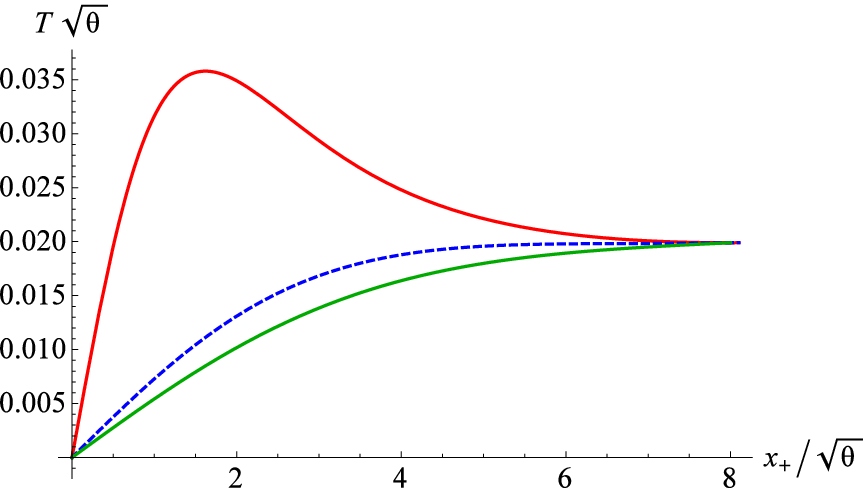}
\caption{The temperature  of the ($1+1$)-dimensional  noncommutative black hole   for $\theta=1$ (red solid curve), for $\theta=\theta_c \approx 7.83$ (blue dashed curve) and for $\theta=13$ (green solid curve). The plot is displayed for  fixed $l=8 \sqrt{\theta}\,$. }
\label{fig:2dNCT}
\end{center}
\end{figure}

\begin{figure}[h!] 
\begin{center}
\includegraphics[width=0.53 \textwidth]{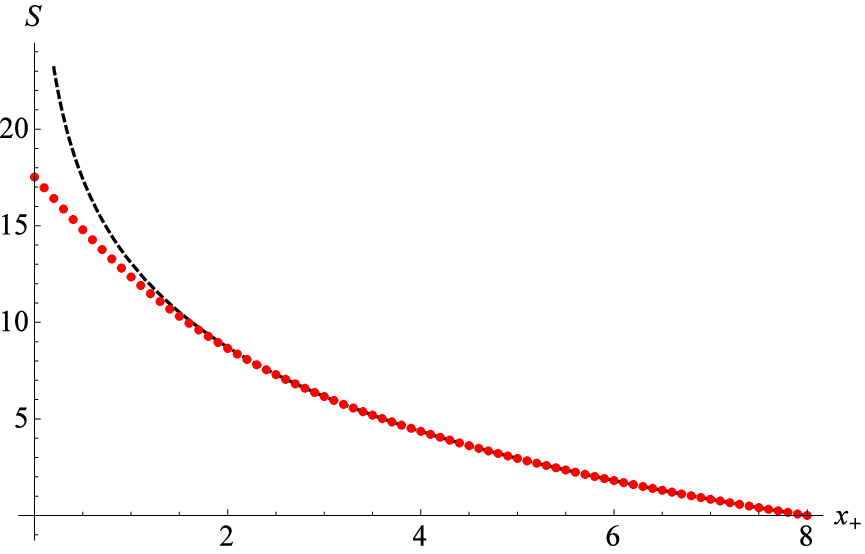}
\caption{The red dots represent the entropy  of a $(1+1)$-dimensional noncommutative black hole  for $\theta=1$ and $l=8 \,$. The black dashed curve represents the singular black hole entropy. }
\label{fig:2dNCS}
\end{center}
\end{figure}

We derive now the desired thermodynamic variables for the noncommutative black hole metric \eqref{2dNCAdS_potential}. From the condition $f(x_+)=0\,$, we express  the black hole mass as
\begin{equation}
M= \frac{\sqrt{\pi} (l^2-x_+^2)}{2l^2 \left( 2\sqrt{\theta} \ e^{-x_+^2/4\theta}  + x_+ \gamma(1/2 , x_+^2/4\theta) \right) } \,.
\end{equation}
We require the mass of the black hole to be positive, and this is satisfied for $x_+ < l\,$, as can be seen in Fig.~\ref{fig:2dNCM}.
The mass decreases with $x_+\,$, as expected in ($1+1$)-dimensions \cite{MaM11}, and vanishes at the AdS radius $l\,$. 
Using next the Hawking formula, we find  the temperature 
\begin{equation} \label{2dNCT}
T= \frac{x_+}{2\pi l^2} \left[1 +  \frac{(l^2-x_+^2) \gamma(1/2,x_+^2/4\theta)}{2 x_+ (2 \sqrt{\theta} e^{-x_+^2/4\theta }+ x_+  \gamma(1/2 , x_+^2/4\theta))} \right] \,.
\end{equation}
For values below the critical $\theta _c\,$, it appears a maximum before running a positive heat capacity, as can be seen in Fig.~\ref{fig:2dNCT}. 
Let us stress something at this point: this temperature appears to behave similarly to that of a Van der Waals gas. However, we must not forget that the mass and the radius are inversely proportional to each other, and this could alter the nature of any potential phase transition. This can be clarified by examining the heat capacities. Based  on the discussion of \cite{FMM15},  we can define the black hole pressure and ``volume", which is actually a line in ($1+1$)-dimensions, by ensuring the validity of the Smarr relation.  This is done by rescaling Newton's constant and taking the limit where the number of dimensions drops to $D \rightarrow 2\,$. This will help us identify the pressure with
\begin{equation} \label{2dNCPV}
P=\frac{1}{8\pi l^2}\,.
\end{equation}
Its conjugate quantity will be
\begin{equation}
V= - \left( \frac{\partial M}{\partial P} \right)_{S,P_{\theta}}=  \frac{4 \pi^{3/2} x_+^2}{2\sqrt{\theta} \ e^{-x_+^2/4\theta} + x_+ \gamma(1/2,x_+^2/4\theta)} \,.
\end{equation}
The negative sign in the definition of $V$ arises from the sign-reversal of $\Lambda\,$, leading us to define the pressure as $P=\frac{\Lambda}{8\pi}$  to ensure its positivity. This adjustment  guarantees that the thermodynamic volume remains positive. In analogy, we can define the noncommutative tension as
\begin{equation}
P_{\theta} = - \frac{1}{8\pi \theta}
\end{equation}
and the noncommutative ``volume" as
\begin{equation}
V_{\theta}= - \left( \frac{\partial M}{\partial P_{\theta}} \right)_{S,P} = \frac{4 (\theta\pi)^{3/2} (l^2-x_+^2)  }{l^2 \left( 2 \sqrt{\theta} e^{-x_+^2/4\theta} + x_+ \gamma(1/2 , x^2/4\theta)\right)^2 } \,
\end{equation}
which also turns out to be positive.
 From the validity of the first law in ($1+1$)-dimensions \cite{FMM15}, i.e., 
\begin{equation}
 \mathrm{d} M=T \mathrm{d} S-V \mathrm{d} P-V_{\theta} \mathrm{d} P_{\theta}\,, 
\end{equation} 
 we find that the black hole entropy is given by
\begin{equation} \label{2dNCS}
S= \int \limits ^{l}_{x_+} \mathrm{d} x \ \frac{2\pi^{3/2} }{2\sqrt{\theta} \ e^{-x^2/4\theta} + x \gamma(1/2,x^2/4\theta)} \,.
\end{equation}
The entropy \eqref{2dNCS} cannot be solved analytically. Therefore, we present a numerical plot of the entropy in Fig.~\ref{fig:2dNCS}.
In the classical limit ($x_+ \gg \sqrt{\theta}$), the expressions of $T,V$ and $S$ align with those of a singular AdS black hole in ($1+1$)-dimensions, as presented in \cite{FMM15}.

\begin{figure}[!h] %
\centering
\subfigure[The specific heat $C_P\,.$ ]{%
\label{fig:2dNCCp}%
\fbox{\includegraphics[height=1.93in]{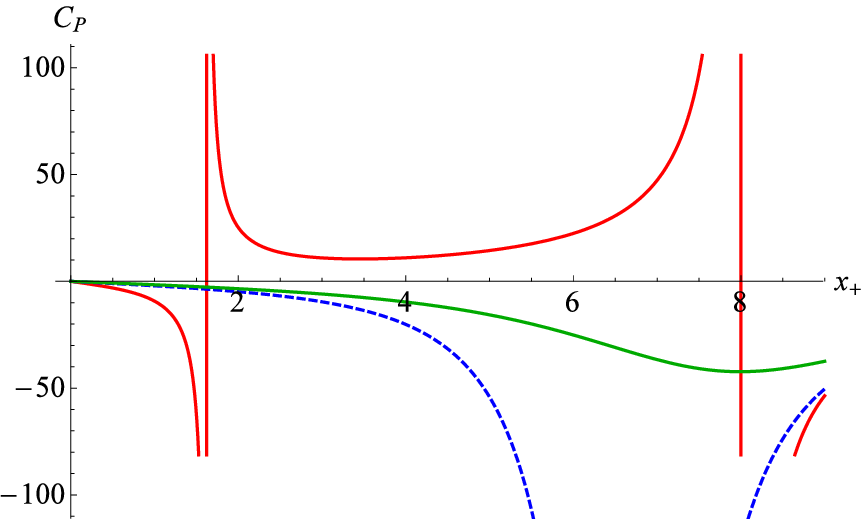}}}%
\qquad
\subfigure[The Gibbs energy $G\,.$]{%
\label{fig:2dNCG}%
\fbox{\includegraphics[height=1.93in]{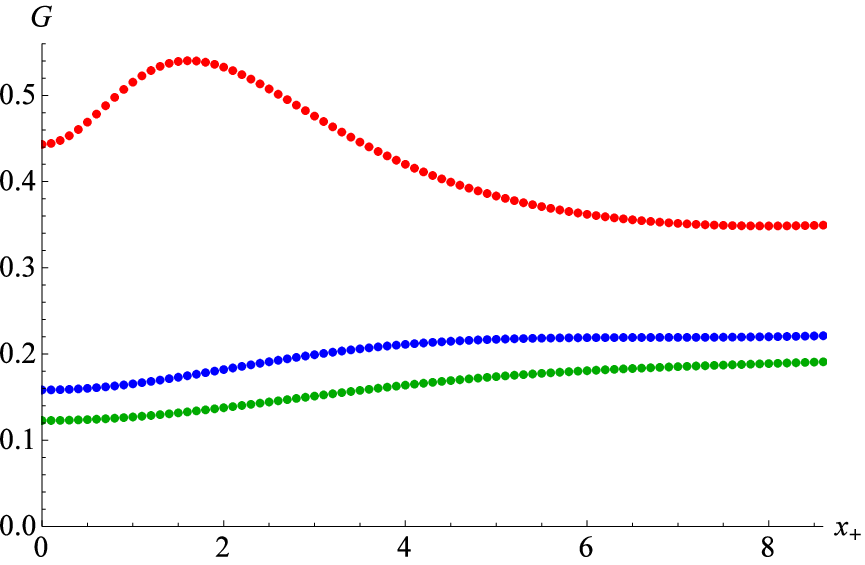}}}%
\caption{The plots of $C_P$ and $G$ for a $(1+1)$-dimensional noncommutative  black hole for $\theta = 1$ (red curve and dots), for $\theta=\theta_c \approx 7.83$ (blue dashed curve and dots) and for $\theta=13$ (green curve and dots). The plots  are displayed  for fixed $l=8 \sqrt{\theta}\,$.}
\label{fig:2dNCCpG}
\end{figure}

Continuing  with the specific heats, we find that $C_V=0\,$, while $C_P$ is calculated to be
\begin{equation}
C_P = \frac{2\pi^{3/2} x_+^2 e^{x_+^2/4\theta} \left[ 4 \theta   + e^{x_+^2 /4\theta} \sqrt{\theta} (l^2+x_+^2) \gamma \right] }{\sqrt{\theta} \left[ 2l^2x_+-2x_+^3+8\theta x_+ + \frac{x_+^2}{\sqrt{\theta}} e^{x_+^2/4\theta} (4\theta+l^2-x_+^2) \right]\gamma + e^{x_+^2/2\theta} x_+ (x_+^2-l^2) \gamma^2 } \,
\end{equation}
where $\gamma \equiv \gamma(1/2,x_+^2/4\theta)\,$. The Gibbs energy can be found through the relation $G=M-T S\,$. The two variables $C_P$ and $G$ are displayed in Fig.~\ref{fig:2dNCCpG}.
For values below the critical  $\theta_c\,$, there is a divergence in $C_P$ at $x_+ = x_{\rm{pt}} \simeq 1.6 \sqrt{\theta}\,$ (red curve  in Fig~\ref{fig:2dNCCp}). For $x< x_{\rm{pt}}$ the black hole is unstable ($C_P<0$) with an increasing $G$, while for $ x_{\rm{pt}} < x < l \,$, the black hole is stable ($C_P>0$) with a decreasing $G$ until it reaches the global minimum  at $r_+ = l=8\sqrt{\theta}$ (red dots  in Fig~\ref{fig:2dNCG}). Thus a phase transition seems to take place between pure AdS radiation and a stable ``large" black hole. However, this transition has opposite thermodynamic behaviour  from the standart Hawking-Page transition \cite{HaP83}, regarding the temperature and the Gibbs energy profile. For this reason, we refer to it as \textit{anti-Hawking-Page  transition}. For values $\theta \geq \theta_c\,,$ the black hole system cannot exist, since it is always thermodynamically unstable with $C_P<0$ (green and blue curve  in Fig~\ref{fig:2dNCCp}). Moreover, the Gibbs energy always increases at this case (green and blue dots  in Fig~\ref{fig:2dNCG}).  Therefore, in contrast to the singular case \cite{FMM15},  where the the black hole system is always thermodynamically stable, the regular black hole appears a phase transition at short distances. Notice that similar behaviour can be extracted by varying $l$ and keeping the scale $\theta$ constant  instead.

\begin{figure}[h!] 
\begin{center}
\includegraphics[width=0.53 \textwidth]{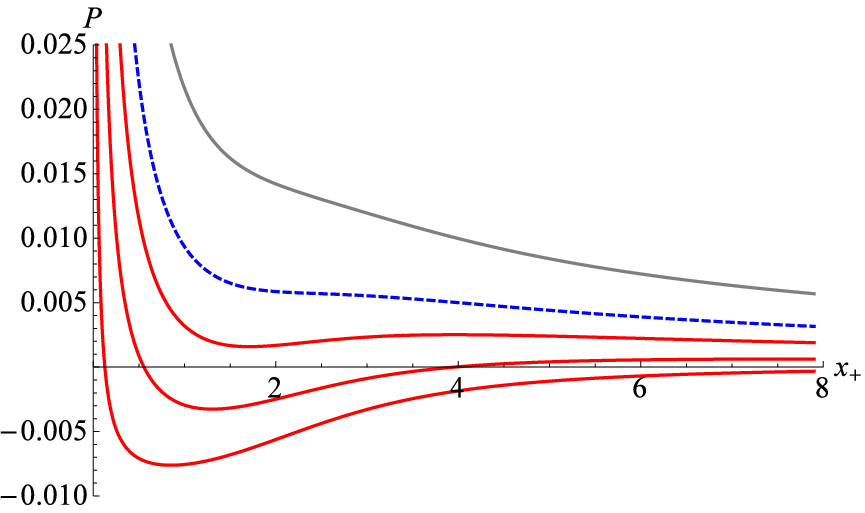}
\caption{The isotherms of the ($1+1$)-dimensional noncommutative  black hole are displayed for $\theta=1\,$. The red solid curves stand for   $T<T_c$, the blue dashed curve stands for the critical temperature, $T=T_c \approx 0.06\,$, and the gray solid curve for $T>T_c\,.$ }
\label{fig:2dNCeos}
\end{center}
\end{figure}

Finally, the black hole equation of state can be found by combining \eqref{2dNCPV} and \eqref{2dNCT}:
\begin{equation}
P= \frac{4\pi T(2 \sqrt{\theta} \  e^{-x_+^2/4\theta} +  x_+ \gamma ) -\gamma }{32 \pi x_+ \sqrt{\theta} \ e^{-x_+^2/4\theta} + 8 \pi x_+^2 \gamma} \,.
\end{equation}
In Fig.~\ref{fig:2dNCeos}, the isotherms of the black hole  appear an unstable branch for values $T<T_c$ (red solid curves in Fig.~\ref{fig:2dNCeos}), corresponding to the radiation/black hole phase transition. For values $T\geq T_c\,$ (blue dashed and gray curves), the black hole cannot support its existence because of local and global  instability.

\section{Conclusions}
\label{concl}

Noncommutative geometry appears to be a powerful tool for exploring the extreme conditions near the center of a black hole. By replacing points, typically described by Dirac delta functions, with very narrow and smooth Gaussian distributions, it offers a more realistic and refined approach to understanding divergent pathologies. After providing a geometric overview of the noncommutative black hole solution in four, three, and two dimensions, we examined its thermodynamic properties within an extended anti-de Sitter phase space. In this framework, the cosmological constant is linked to the thermodynamic pressure of the system.

In conclusion, the ($3+1$)-dimensional noncommutative AdS  black hole exhibits chemical characteristics similar to the Van der Waals gas. Its equation of state, along with its heat capacity and Gibbs free energy, suggests a phase transition between small/large stable black holes, analogous to the liquid/gas transition, for pressures below the critical value. Moreover, we treated the minimal cut-off length $\sqrt{\theta}$ as a thermodynamic parameter, associated with the noncommutative tension, and derived the  first law of thermodynamics. The entropy derived from the first law is given by the area law up to some quantum corrections. 

As for the lower dimensional cases, the ($2+1$)-dimensional NCBTZ black hole provides a framework for both global and local stability, without exhibiting any chemical aspects, similar to its singular counterpart. In contrast, the ($1+1$)-dimensional noncommutative case reveals a phase transition between AdS radiation and a stable black hole, which we refer to as the anti-Hawking-Page transition.

It seems that noncommutative geometry naturally extends the classical picture by fixing the  infinite behaviours that appear at very small scales, both in the geometry and in the thermodynamics of black holes. By introducing a cut-off length scale, it smooths out the singularities that usually appear at the black hole center, making the physical quantities well-defined and finite. This leads to new and interesting features, such as stable black hole remnants and richer phase transitions, which are not seen in the  classical scenario. These results suggest that noncommutative geometry captures important effects of quantum gravity and offers a promising way to better understand how singularities might be resolved. Overall, these results highlight the power of noncommutative geometry to address long-standing issues in black hole physics and emphasize its potential role in advancing our understanding of quantum gravity effects.


\end{document}